\documentclass[10pt]{article}
\usepackage[margin=1.5cm]{geometry}
\usepackage{url}
\begin{document}
\title{\Large{Traversing News with Ant Colony Optimisation and Negative Pheromones}\thanks{accepted as preprint for oral presentation at ECCS'14 in Lucca, Italy}}
\author{\large David M.S. Rodrigues and Vitorino Ramos {\small[\url{david.rodrigues@open.ac.uk}, \url{vitorino.ramos@ist.utl.pt}]}}
\date{}
\maketitle

The past decade has seen the rapid development of the online newsroom. News published online are the main outlet of news surpassing traditional printed newspapers. This poses challenges to the production and to the consumption of those news. With those many sources of information available it is important to find ways to cluster and organise the documents if one wants to understand this new system.

Traditional approaches to the problem of clustering documents usually embed the documents in a suitable \textit{similarity space}. Previous studies have reported on the impact of the similarity measures used for clustering of textual corpora~\cite{strehl2000impact}. These similarity measures usually are calculated for bag of words representations of the documents. This makes the final document-word matrix high dimensional. Feature vectors with more than 10,000 dimensions are common and algorithms have severe problems with the high dimensionality of the data.

A novel bio inspired approach to the problem of traversing the news is presented. It finds Hamiltonian cycles over documents published by the newspaper \textit{The Guardian}. A \textit{Second Order Swarm Intelligence} algorithm based on Ant Colony Optimisation was developed~\cite{Rodrigues:2011fj,Ramos:2013qy} that uses a negative pheromone to mark unrewarding paths with a ``no-entry'' signal. This approach  follows recent findings of negative pheromone usage in real ants~\cite{Robinson:2007}. 

In this case study the corpus of data is represented as a bipartite relation between documents and keywords entered by the journalists to characterise the news. A new similarity measure between documents is presented based on the \textit{Q}-analysis description~\cite{Atkin:1974,Johnson:1983c,rodrigues:2013aa} of the simplicial complex formed between documents and keywords.  The eccentricity between documents (two simplicies) is then used  as a novel measure of similarity between documents. 

The results prove that the \textit{Second Order Swarm Intelligence} algorithm performs better in benchmark problems of the travelling salesman problem, with faster convergence and optimal results. The addition of the negative pheromone as a non-entry signal clearly improved the quality of the solutions. The application of the algorithm to the corpus of news of \textit{The Guardian} creates a coherent  navigation system among the news. This allows the users to navigate the news published during a certain period of time in a semantic sequence instead of a time sequence.  

This work as broader application as it can be applied to many cases where the data is mapped to bipartite relations (e.g. protein expressions in cells, sentiment analysis, brand awareness in social media, routing problems), as it highlights the connectivity of the underlying complex system.

\end{document}